\documentclass[aps,prb,twocolumn,showpacs,superscriptaddress]{revtex4}

\usepackage{graphicx}
\usepackage{amsmath}

\usepackage[colorlinks,linkcolor=blue,anchorcolor=blue,citecolor=blue]{hyperref}

\begin{document}

\title{Spin dynamics of the spin chain antiferromagnet RbFeS$_2$}

\author{Lisi Li}
\affiliation{Center for Neutron Science and Technology, School of Physics, Sun Yat-Sen University, Guangzhou, 510275, China }
\author{Liangliang Zheng}
\affiliation{Center for Neutron Science and Technology, School of Physics, Sun Yat-Sen University, Guangzhou, 510275, China }
\author{Benjamin A. Frandsen}
\affiliation{Materials Science Division, Lawrence Berkeley National Laboratory, Berkeley, California 94720, USA}
\affiliation{Department of Physics and Astronomy, Brigham Young University, Provo, Utah 84602, USA}
\author{Andrew D. Christianson}
\affiliation{Materials Science and Technology Division, Oak Ridge National Laboratory, Oak Ridge, Tennessee 37831, USA}
\author{Dao-Xin Yao}
\affiliation{Center for Neutron Science and Technology, School of Physics, Sun Yat-Sen University, Guangzhou, 510275, China }
\author{Meng Wang}
\email{wangmeng5@mail.sysu.edu.cn}
\affiliation{Center for Neutron Science and Technology, School of Physics, Sun Yat-Sen University, Guangzhou, 510275, China }
\author{Robert J. Birgeneau}
\affiliation{Department of Physics, University of California, Berkeley, California 94720, USA }
\affiliation{Materials Science Division, Lawrence Berkeley National Laboratory, Berkeley, California 94720, USA}

\begin{abstract}
We report transport and inelastic neutron scattering studies on electronic properties and spin dynamics of the quasi-one-dimensional spin chain antiferromagnet RbFeS$_2$. An antiferromagnetic phase transition at $T_N\approx195$ K and dispersive spin waves with a spin gap of 5 meV are observed. By modeling the spin excitation spectra using linear spin wave theory, intra and inter-chain exchange interactions are found to be $SJ_1=100(5)$ meV and $SJ_3=0.9(3)$ meV, respectively, together with a small single-ion anisotropy of $SD_{zz}=0.04(1)$ meV. Comparison with previous results for other materials in the same class of Fe$^{3+}$ spin chain systems reveals that although the magnetic order sizes show significant variation from 1.8 to 3.0$\mu_B$ within the family of materials, the exchange interactions $SJ$ are nevertheless quite similar, analogous to the iron pnictide superconductors where both localized and delocalized electrons contribute to the spin dynamics.
\end{abstract}

\maketitle

\section{Introduction}

The discovery of iron-based superconductors has attracted significant scientific interests\cite{Kamihara2008,Stewart2011,Fang2013}. Most iron-based superconductors crystallize in quasi-two-dimensional (2D) layered structures consisting of Fe$Pn_4$ or Fe$Ch_4$ ($Pn$ = pnictogens, $Ch$ = chalcogens) tetrahedra\cite{Ryu2012,Guo2010,Mazin2008,Yan2012,Fang2011}. More recently, the observation of pressure-induced superconductivity in the quasi-one-dimensional (1D) spin-ladder compound BaFe$_2$S$_3$, which also consists of FeS$_4$ tetrahedra, has drawn attention to the 1D iron-chalcogenide materials\cite{Takahashi2015,Yamauchi2015}. Adding to this interest in 1D materials is a recent investigation of the 1D spin-chain compound TlFeSe$_2$ under pressure, which is metallized above 2 GPa and may support superconductivity above 30 GPa\cite{Liu2020}. More broadly speaking, iron-based superconductors and related materials exemplify the close relationship between magnetism and superconductivity in structures comprised of 2D layers, 1D ladders, and 1D chains. Nevertheless, a complete understanding of the impact of structure on magnetism and superconductivity remains elusive, motivating continued study in this field.

\begin{figure}
\centering
\includegraphics[width=6.5 cm]{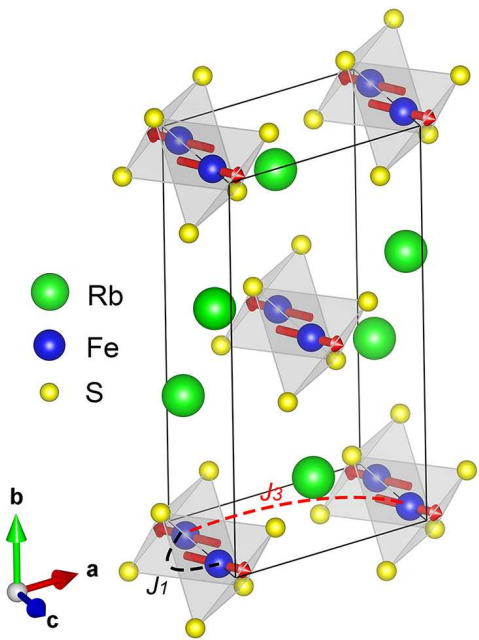}\\
\caption{Crystal and magnetic structure of RbFeS$_2$. Rb, Fe, and S atoms are drawn as green, blue, and yellow spheres, respectively. The exchange interactions $J_1$ and $J_3$ are marked with black and red dashed lines, respectively. The magnetic structure adapts to the published results\cite{Bronger1987}, the red arrows represent the magnetic moment direction.}
\label{fig1}
\end{figure}

In this context, the ternary metal chalcogenides $A$Fe$X_2$ ($A$=alkali metal, Tl; $X$=S,Se) are interesting materials because they host linear spin chains formed by edge-sharing $^1_\infty$[Fe$X_{4/2}$]$^-$ tetrahedra along the chain axis, as shown in Fig. \ref{fig1} (a)\cite{Bronger1987,Asgerov2015,Bronger1968}. As previously reported, KFeS$_2$, RbFeS$_2$, KFeSe$_2$, and RbFeSe$_2$ crystallize in the monoclinic space group $C/2c$, while TlFeS$_2$ and TlFeSe$_2$ crystallize in the monoclinic space group $C/2m$\cite{Nishi1979,TOMKOWICZA1980,Asgerov2014}. CsFeS$_2$ is somewhat different, crystallizing in an orthorhombic space group and undergoing a structural transition from $Immm$ to $P \bar{1}$ upon cooling\cite{Bronger1968,Tiwary1997,Welz1997}.

\begin{table*}
\label{table1}
\caption{Crystal and magnetic information for the 1D spin chain compound $A$Fe$X_2$ ($A$ = K, Rb, Tl; $X$ = S, Se). $E^{L}_{t}$ and ${\Delta}_{s}$ represent the zone-boundary energy and the spin gap of the spin wave along the chain direction.}
\setlength{\tabcolsep}{2mm}
\begin{tabular}{cccccccc}
\hline \hline
Compound  &Space  &$T_\text{N}$	&$d$ ({\AA}) &Moment 	&$\mu_\textrm{order}$  &Spin wave     &Ref.\\
~    &group  &(K) &(Fe-Fe) &orientation &($\mu_B$) & & \\
\hline
KFeS$_2$   &$C/2c$	 &250.0(5) &2.70 &13.6$^\circ$ from chain &2.43(3)  &$E^{L}_{t}=221(4)$ meV &[\cite{Bronger1987,Nishi1979,TOMKOWICZA1980,Funahashi1983,Welz1992a}] \\
~ &  &  &  &  & &${\Delta}_{s} \approx$ 4.5 meV &~ \\ \hline
RbFeS$_2$  &$C/2c$	&188(1) &2.71 &20$^\circ$ from chain  &1.8(3) & &[\cite{Bronger1987,Bronger1997}]\\
&~       &     &   & & &$E^{L}_{t}\approx203$meV &\\
~ &   &$\approx195$ &  &close to the chain  &        &${\Delta}_{s}\approx5$ meV &[this work] \\ \hline
KFeSe$_2$  &$C/2c$    &310(1) &2.81  &$\bot$ chain  &3.0(2)    &-  &[\cite{Bronger1987}]  \\ \hline
RbFeSe$_2$ &$C/2c$   &249(1) &2.83  &$\bot$ chain  &2.66(5)  &-  &[\cite{Bronger1987,Seidov2016,Kiiamov2018}]  \\ \hline
TlFeS$_2$  &$C/2m$	 &196 &2.65 &$\bot$ chain  &1.85(6)  &$E^{L}_{t}\approx260$ meV   &[\cite{Welz1989,Welz1992,Welz1996}]\\
~  & & & & & &${\Delta}_{s} \approx4.3$ meV  &~\\ \hline
TlFeSe$_2$ &$C/2m$  &290 &2.74 &$\bot$ chain  &2.1(2)   &-  &[\cite{Seidov2001,Asgerov2014,Asgerov2015}] \\ \hline
\hline \hline
\end{tabular}
\end{table*}

These spin chain compounds host trivalent iron ions with a 3$d^5$ electron configuration. The chains of Fe$^{3+}$ spins all exhibit antiferromagnetic (AFM) order at low temperature, with the moments aligning perpendicularly to the spin chain axis except in the case of KFeS$_2$ and RbFeS$_2$, in which the moments deviate from the spin chain axis by relatively small angles (see Table \ref{table1} \cite{Bronger1987}). Interestingly, the observed ordered moment sizes are in the range $1.8\sim3\mu_B$, smaller than the expected size of 5$\mu_B$ for Fe$^{3+}$ spins. Furthermore, a direct measurement of the bulk electronic structure by $2p$ core-level hard x-ray photoemission spectroscopy and calculation density of states in TlFeS$_2$ and TlFeSe$_2$ suggest that the competition between the delocalized and localized characters of Fe 3d electrons is important to understand the electronic structures of both compounds\cite{Mimura2013}. For these reasons, it has been suggested that the Fe$^{3+}$ 3$d^5$ electrons exhibit considerable delocalization and a spin $S=3/2$, even though these compounds are showing semiconducting behaviors at ambient pressure. This delocalization may be related to the short nearest neighbor (NN) intrachain atomic distance of Fe-Fe, which approaches the metallic bond distance of Fe. This could result in Fe-Fe covalency in addition to Fe-S covalency, thus reducing the magnetic moment. The lack of metallic conductance in the AFe$X_2$ system is likely related to the 1D character of the structure.

Given the short NN Fe-Fe bond, a strong intrachain magnetic exchange interaction should be expected. For TlFeS$_2$ (space group $C/2c$), inelastic neutron scattering (INS) measurements determined that the NN AFM intrachain exchange interaction $SJ$ is $\sim$ 65 meV\cite{Welz1992,Welz1996}. Additionally, the interchain-intrachain exchange ratio is of order 10$^{-3}$, confirming a strong 1D behavior in the magnetic interactions. However, KFeS$_2$ (space group $C/2m$) is reported to have a significantly smaller NN AFM intrachain exchange interaction $SJ$=25(1) meV \cite{DeBiasi1978,Funahashi1983,Welz1992a}. The result was deduced only from the low-energy spin excitations and may not be accurate \cite{Funahashi1983}. In a later INS experiment with higher incident energies, the spin waves were observed to extend to 221(4) meV at the zone boundary. However, the NN intrachain exchange interaction was not extracted from the data\cite{Welz1992a}. Considering this uncertainty in the exchange parameters for KFeS$_2$ (and the complete lack of such information for RbFeS$_2$, which shares the $C/2m$ structure with KFeS$_2$), it would be valuable to clarify the exchange parameters for the $C/2m$ systems and compare them to the representative $C/2c$ compound TlFeS$_2$. Such a comparison would further elucidate the relationship between structure and magnetism in low-dimensional iron-based systems.

In this paper, we report transport and inelastic neutron scattering (INS) measurements on RbFeS$_2$. We find an AFM phase transition takes place at $T_N\approx195$ K with magnetic moment oriented close to the chain direction. In the INS spectra, we observe two branches of spin waves along the $H$ and $L$ directions and a spin gap located at the Bragg peak positions. The energies of the band-top and spin gap are close to those of the isostructural compound KFeS$_2$, which has a moment size of $\sim2.43(3)\mu_B$ for Fe$^{3+}$. By modeling these spectra using a linear spin wave theory, we find that the spectra can be fully reproduced with an intrachain exchange of $SJ_1 = 100(5)$ meV, an interchain exchange of $SJ_3 = 0.9(3)$ meV, and a small single-ion anisotropy of $SD_{zz}=0.04(1)$ meV. These results demonstrate that although a wide variety of magnetic structures, ordered moment sizes, and electron itinerancy are observed among the 1D spin chain compounds, the spin excitations are nevertheless quite similar across the family of compounds. This suggests an interplay between localized and delocalized magnetism in the 3$d$ bands of Fe$^{3+}$ reminiscent of the 2D iron-based superconductors.

\begin{figure} [t]
\centering
\includegraphics[width= 8 cm]{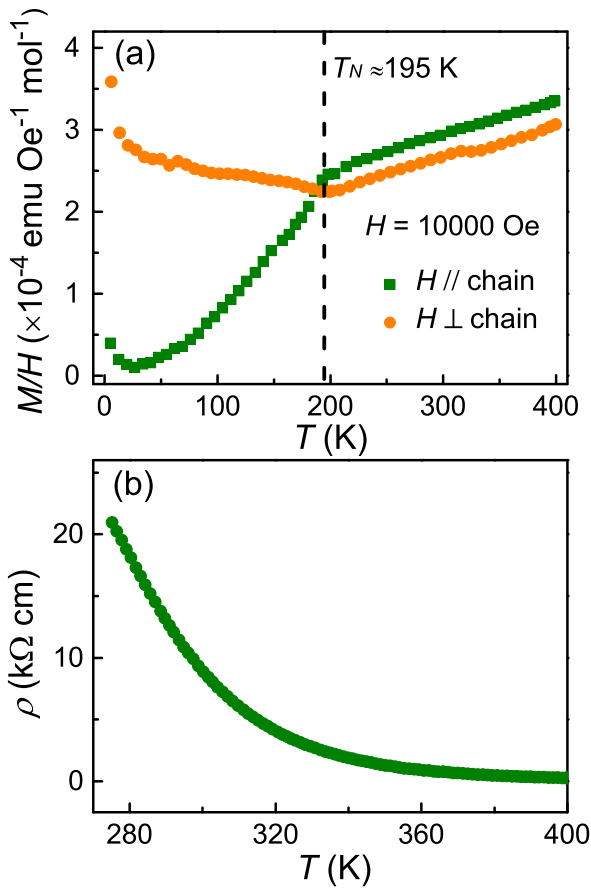}\\
\caption{(a) Temperature dependence of the susceptibility of RbFeS$_2$ measured under magnetic fields applied parallel and perpendicular to the chain direction. (b) Temperature dependence of the resistivity measured along the chain direction. }
\label{fig2}
\end{figure}

\begin{figure*}
\includegraphics[width=18cm]{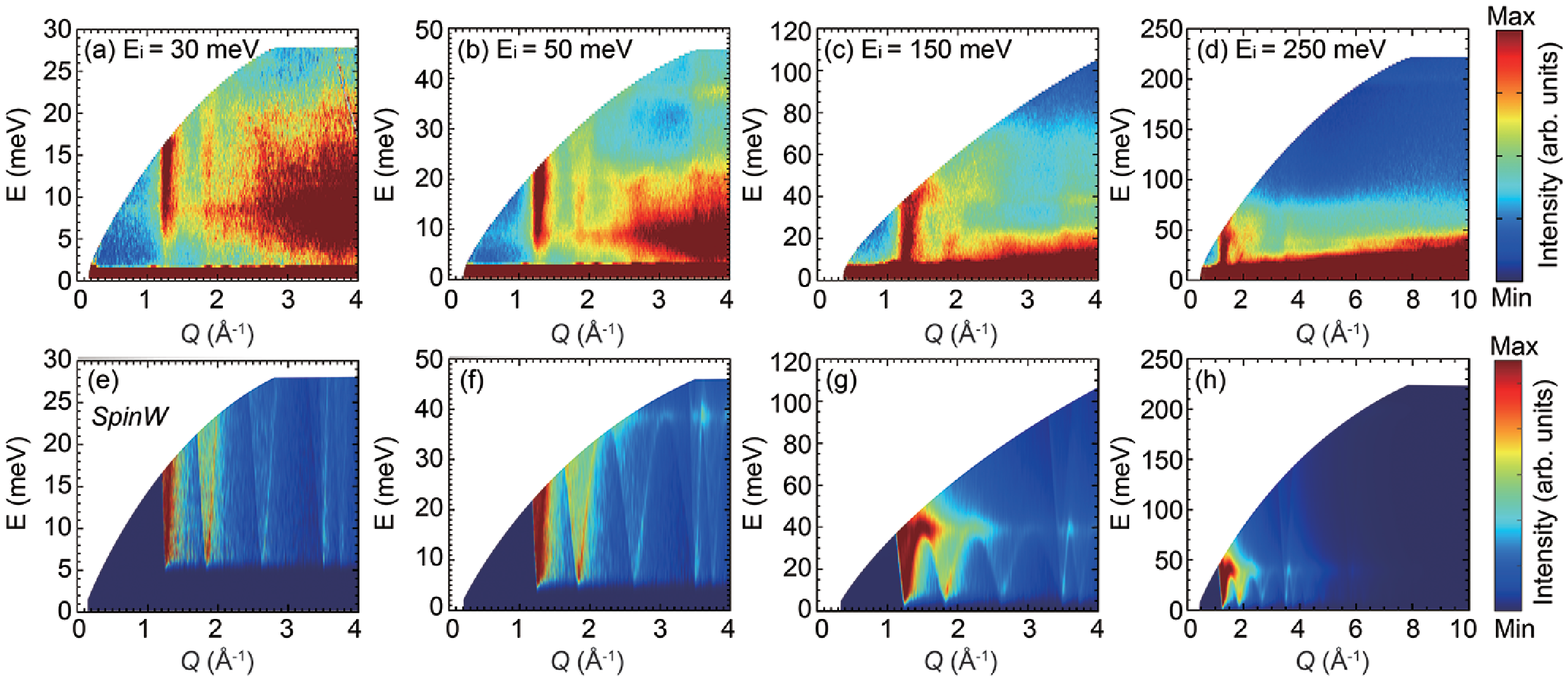}\\
\caption{INS spectra $S(Q,\omega)$ of RbFeS$_2$ at 4 K with $E_i$= (a) 30, (b) 50, (c) 150, and (d) 250 meV. $SpinW$ simulated spectra $S(Q,\omega)$ with the exchange parameters described in the text. The instrumental resolutions of (e) 1.2, (f) 2.2, (g) 6.2, and (h) 11.4 meV have been convoluted in the simulation for comparing to the INS spectra with $E_i$ = 30, 50, 150, and 250 meV, respectively. The color represents intensities. }
\label{fig3}
\end{figure*}

\section{Experimental details}

Single crystal samples of RbFeS$_2$ were grown using the Bridgman method. The sintering procedure is identical to that of Rb$_{0.8}$Fe$_{1.5}$S$_2$ we grew previously\cite{wang2014}. The obtained single crystals are needle-like in shape, consistent with the 1D structure. DC susceptibility and resistivity data were collected on a commercial physical property measurement system (PPMS, Quantum Design). The needle-like single crystals are difficult to align, we thus ground 3 g of single crystals into powder for neutron scattering experiments. The INS experiment was performed on the ARCS time-of-flight chopper spectrometer at the Spallation Neutron Source, ORNL\cite{Abernathy2012}. The powder sample was sealed in an aluminum can and loaded into a He top-loading refrigerator. The sample was measured with incident energies of $E_{i}$ = 30, 50, 150, and 250 meV at 4 K with the corresponding energy resolutions $\Delta E$ = 1.2, 2.2, 6.2, and 11.4 meV, respectively, as determined by the full width at half maximum (FWHM) of the energy cuts at $E=0$ meV. The data reduction for the INS data was performed using the $DAVE$ software\cite{Azuah2009}. Linear spin wave theory was employed to simulate the INS spectra using the $SpinW$ software\cite{Toth2015}.\\

\section{Results}

Figure \ref{fig2} (a) shows the dc susceptibility measured under magnetic fields applied parallel ($\chi_\|$) and perpendicular ($\chi_\bot$) to the chain direction. Both the $\chi_\|$ and $\chi_\bot$ exhibit a slop change at $T_N\approx195$ K, revealing an AFM phase transition. Below $T_N$, the $\chi_\bot$ is larger than $\chi_\|$ in magnitude, suggesting that the magnetic easy axis is close to the chain direction. The results are in agreement with the previous results that the AFM phase transition of RbFeS$_2$ is at $T_N=188(1)$ K and the magnetic moment arranges in the $ac$ plane with an angle of $20^\circ$ to the $c$ axis as shown in Fig. \ref{fig1}\cite{Bronger1987}. The upturns at low temperature reveal the existence of paramagnetic impurities. Above $T_N$, the $\chi_\|$ and $\chi_\bot$ increase linearly with increasing temperature, resembling that of TlFeS$_2$, TlFeSe$_2$, and RbFeSe$_2$ \cite{Seidov2001,Seidov2016}.
The linear increase of susceptibility with the increasing temperature also appears in some quasi two-dimensional metallic layered iron pnictides above its AFM transition\cite{Stewart2011}, such as in Ba(Fe$_{1-x}$Co$_x$)$_2$As$_2$ ($x=0-0.2$)\cite{Wang2009,Wang2009b}, Ca(Fe$_{1-x}$Co$_x$)$_2$As$_2$ ($x=0-0.2$)\cite{Harnagea2011}, and LaFeAsO$_{1-x}$F$_x$ ($x=0-0.15$)\cite{Klingeler2010}.
The linear susceptibility behavior could be attributed to the antiferromagnetic correlation of local moment in a strong coupling description \cite{Kou2009,Wang2009} or the antiferromagnetic spin fluctuations of itinerant electron \cite{Zhang2009, Korshunov2009}.
Figure \ref{fig2} (b) shows the temperature dependence of the resistivity down to 275 K, indicating a semiconducting behavior. Below this temperature, the resistance is out of the limitation of our instrument.
The semiconducting behavior was also observed in the 1D system such as KFeS$_2$, TlFeS$_2$, TlFeSe$_2$, and RbFeSe$_2$\cite{Nishioka1995,Seidov2001,Seidov2016}, where it is ascribed to the fiber-like morphology of the 1D structure which contains defects and breaks in the sample.

In Figs \ref{fig3} (a-d), we present INS spectra obtained from the powder sample at 4 K using incident neutron energies of $E_{i}=30$, 50, 150, and 250 meV, respectively. We observe intense excitations at $Q=1.25$ and $Q=1.82$ {\AA}$^{-1}$, along with a spin gap of $\sim5$ meV. Additionally, excitations stemming from $Q=3.5$ {\AA}$^{-1}$ with much weaker intensities appear at $E_{i}=150$ and $E_{i}=250$ meV spectra, as shown in Figs. \ref{fig3} (c) and (d). The intensities of the excitations decrease with increasing $Q$, consistent with a magnetic origin. The three $Q$ values from which we observe excitations are correspond to the AFM wave vectors $(H, K, L)=(0, 0, 1), (1, 0, 1),$ and $(3, 0, 1)$, respectively, demonstrating that these are spin wave excitations out of the magnetic ground state. Here, $(H, K, L)$ are Miller indices for the momentum transfer $Q=2\pi\sqrt{(\frac{H}{a\sin\beta})^2+(\frac{K}{b})^2+(\frac{L}{c\sin\beta})^2-\frac{2HL\cos\beta}{ac(\sin\beta)^2}}$, where the lattice parameters are $a=7.162(7), b=11.566(7),$ and $c=5.453(5)$ {\AA}, and $\beta=112.75(7)^\circ$ obtained from refining the energy cuts of $E_{i}=30$ meV spectrum at $E=0$ meV at 4 K. The flat excitations below 30 meV that increase in intensity with $Q$ correspond to phonon from the sample and the thin aluminum can.

\begin{figure}
\centering
\includegraphics[width=8.5cm]{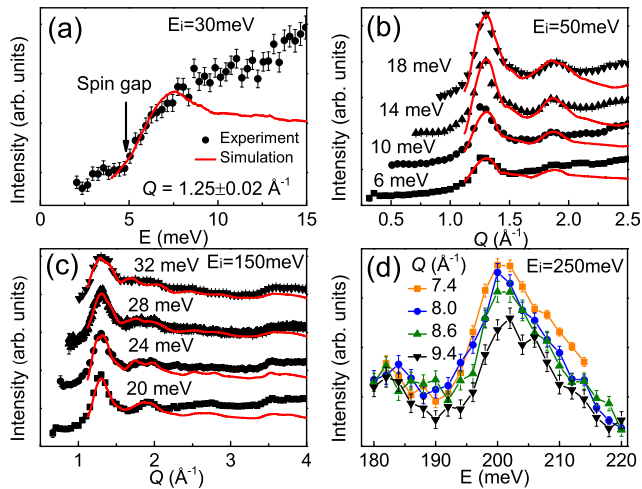}
\caption{(a) Constant $Q$ cut between $1.23< Q <1.27$ \AA$^{-1}$ for $E_{i}=30$ meV. (b) Constant energy cuts at $E$ = 6, 10, 14, and 18 meV integrated within $E\pm2$ meV for $E_{i}=50$ meV. (c) Constant energy cuts at $E$ = 20, 24, 28, and 32 meV integrated within $E\pm2$ meV with $E_{i}=150$ meV. The black symbols are experimental data while the red solid lines depict simulated results. (d) Constant $Q$ cuts at $Q$ = 7.4, 8, 8.6, and 9.4 {\AA}$^{-1}$ integrated within $Q\pm0.2$ {\AA}$^{-1}$ with $E_{i}=250$ meV. }
\label{fig4}
\end{figure}

Taking a more quantitative view of the spin gap and spin wave dispersions, we show constant $Q$ and $E$ cuts in Fig. \ref{fig4}. Panel (a) displays the constant $Q$ cut integrated over $Q=1.25\pm0.2$ {\AA}$^{-1}$ with $E_i=30$ meV. The abrupt increase at $\Delta E\approx5$ meV confirms a spin gap of $\Delta_s\approx5$ meV. Figures \ref{fig4} (b) and (c) show constant $E$ cuts with $E_i$ = 50 and 150 meV, respectively. The excitation near $Q=1.82$ {\AA}$^{-1}$ shows clear dispersion with increasing energy transfer. The excitation corresponding to $(H, K, L)=(3, 0, 1)$ can also be recognized around $Q=3.5$ {\AA}$^{-1}$ in Fig. \ref{fig4} (c), also showing a continuous evolution with increasing energy transfer. Constant $Q$ cuts at $Q=7.4, 8.0, 8.6$ and 9.4 {\AA}$^{-1}$ are displayed for the energy range $180-220$ meV ($E_i=250$ meV) in Fig. 4 (d). The high energy excitations located around $E\approx 203$ meV are much higher than the cut-off energy of phonon. Instead, these high-energy excitations correspond to the band-top of the spin waves at the zone-boundary along the $L$ direction. The INS spectra are comparable with the spin waves of KFeS$_2$ and TlFeS$_2$ as shown in Table \ref{table1}\cite{Welz1992a,Welz1992,Welz1996}.

Having measured the spin wave dispersion, we now turn to modeling the INS spectra using linear spin wave theory with the following Heisenberg Hamiltonian:

\begin{equation}
  \hat{H}=\sum_{i,j}{J_{i,j}}\emph{\textbf{S}}_\emph{i}\cdot \emph{\textbf{S}}_\emph{j}-\it{D_{zz}}\sum_{i,z}S_{i,z}^\textrm{2},
  \label{eq1}
\end{equation}

where $J_{i,j}$ includes the NN intrachain exchange interaction $J_1$ and NN interchain exchange interaction $J_3$ as marked in Fig. \ref{fig1}. $D_{zz}$ is the single ion anisotropy term. Since we observe the spin waves originating only from the AFM wave vectors $(H,0,L)$ in our INS spectra, we only take into account the NN exchange interactions along the $H$ and $L$ directions. By solving Eq. \ref{eq1} using the linear spin wave approximation, the dispersion relations could be written as:

\begin{equation}
\begin{aligned}
&E=\sqrt{A_k^2-B_k^2},\\
&A_k=S(2J_1+2J_3+D_{zz}),\\
&B_k=S(2J_1\cos(\pi L)+2J_3\cos(2\pi H+\pi L)),\\
\label{eq2}
\end{aligned}
\end{equation}

From Eq. \ref{eq2}, the extremes of the spin waves can be extracted. The spin gap $\Delta_{s}$ and the top of the acoustic spin wave along the $H$ ($E^{H}_{t}$) and $L$ directions ($E^{L}_{t}$) are obtained as follows:

\begin{equation}\label{eq3}
\begin{aligned}
&\Delta_{s}\approx S\sqrt{D_{zz}(4J_1+4J_3+D_{zz})},\\
&E^{H}_{t}\approx S\sqrt{16J_1J_3+D_{zz}(4J_1+4J_3+D_{zz})},\\
&E^{L}_{t}\approx S(2J_1+2J_3+D_{zz}).\\
\end{aligned}
\end{equation}

Based on the experimental data, we determine the values for these extremes as $\Delta_s\approx5, E_{t}^{H}\approx40,$ and $E_{t}^{L}\approx203$ meV. From this, we determine the products of the spin $S$ and the magnetic exchange interactions to be $SJ_1=100(5), SJ_3=0.9(3)$, and $SD_{zz}=0.04(1)$ meV. The errors are estimated by allowing 5$\%$ uncertainty of the experimental determined extremes in Eq. \ref{eq3}. The small $SD_{zz}=0.04(1)$ meV of RbFeS$_2$ yields an isotropic magnetic behavior, which can be further verified by the Land\'{e} $g$ factor of 2.00064 measured with electron-spin-resonance (ESR). The $g$ factor closing to the spin only value of $g=2$ suggests a small residual orbital moment\cite{DeBiasi1981,DeBiasi1978}.

\begin{figure}[t]
\centering
\includegraphics[width=8.5cm]{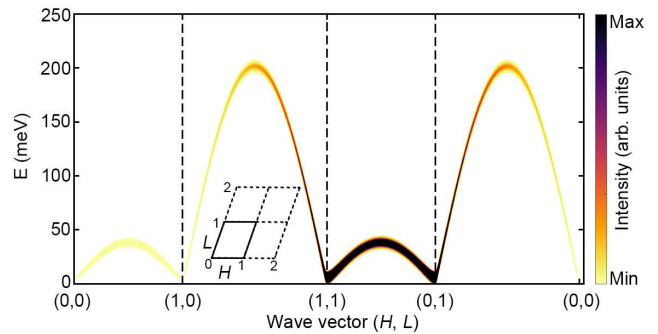}
\caption{Spin waves for single crystals of  RbFeS$_2$ simulated using the $SpinW$ software along the high-symmetry directions in the [$H, L$] Brillouin zone with the path depicted in the inset. An instrumental resolution of 5 meV is convoluted for visualization. The color represents intensities.}
\label{fig5}
\end{figure}

The $SpinW$ software is employed to simulate the spherically averaged spin wave spectra based on the above-determined exchange interactions. The results after convolution with the experimental resolution function are plotted in Figs. \ref{fig3} (e)-(h). The simulated spectra match well with the experimental data. The band-top of the spin waves at the zone-boundary along the $L$ direction is invisible in Fig. \ref{fig3} (h) because the intensities of the spin waves are greatly reduced due to the effect of the magnetic form factor of Fe$^{3+}$. We also perform identical constant energy cuts for the simulated spectra and plot them with the experimental data together in Figs. \ref{fig4} (b) and (c). The energy cuts from the simulated spectra are in good agreement with the experimental results, demonstrating a high accuracy of the determination of the products $SJ$s. In the high $Q$ region, there are discrepancies between the simulated and experimental results, more obvious at energy transfer below 30 meV, which are attribute to the strong phonon intensities in the experimental spectra. To visualize the spin waves more clearly, we plot in Fig. \ref{fig5} the dispersions along high-symmetry directions in the [$H,L$] Brillouin zone using the obtained exchange interactions. The dispersion relations of two acoustic spin wave branches extending to $\sim40$ and $\sim203$ meV along the $H$ and $L$ directions are in good agreement with the experimental observations.

\section{Discussion}

By utilizing the cutting edge time-of-flight neutron scattering spectrometer, two branches of the acoustic spin waves of RbFeS$_2$ have been observed on a powder sample. The spin waves of RbFeS$_2$ and the fitted magnetic exchange interactions resemble the 1D analogs KFeS$_2$ and TlFeS$_2$\cite{Welz1992a,Welz1992}. The ordered moment size is reduced on the sample we measured compared with the expectation for the localized Fe$^{3+}$ $3d$ electrons. The reduction may result from the delocalization of the $3d$ electrons of Fe$^{3+}$ and the quantum fluctuations because of intrinsic nature of the 1D spin chain\cite{Jongh1974,Welz1996}.

Owing to the 1D nature, the single crystals are easily disassembled into thin fibers that result in poor electric conductivity\cite{Seidov2001}. However, the $2p$ core-level Hard x-ray photoemission spectra of TlFeS$_2$ and TlFeSe$_2$\cite{Mimura2013} reveal the delocalization of the Fe$^{3+}$ $3d$ electrons.
Based on  the reduced magnetic moment of 1.8(3) $\mu_B$ and the similarity to TlFeS$_2$ and TlFeSe$_2$, the delocalization of the 3$d$ electrons of Fe$^{3+}$ may exist in RbFeS$_2$ as well. In this scenario, the linear magnetic susceptibility of the ternary metal chalcogenides $A$Fe$X_2$ ($A$=alkali metal, Tl; $X$=S,Se) could be attributed to the antiferromagnetic spin correlations of local moments or spin fluctuations of itinerant electrons, analogous to the iron-based superconductors with the Fe$^{2+}$ 3$d$ electrons \cite{Kou2009,Wang2009,Zhang2009, Korshunov2009}.
The robust spin excitations in KFeS$_2$, RbFeS$_2$, and TlFeS$_2$ in spite of the varied magnetic orders and moment sizes suggest both local moments and delocalized electrons contribute to spin dynamics.
This resembles the spin dynamics of the 2D iron-based superconductors, where local moments and itinerant electrons couple and high energy spin excitations are robust against electron or hole doping \cite{Wang2013, Tam2015}.
This inspires more researches on exploring interesting physics, such as insulator-metal transition and superconductivity on the 1D spin chain system.
We note the spin wave spectra may deviate from the Heisenberg Hamiltonian due to the existence of the delocalized electrons, which call for further studies on the spin waves of single crystal samples.

\section{Conclusions}

In summary, we have measured the magnetic transport and spin waves of the AFM spin chain compound RbFeS$_2$. The susceptibility measurements yield an AFM phase transition at $T_N\approx195$ K with magnetic moment oriented close to the chain direction. The reduced magnetic moment and the similarity of RbFeS$_2$ to TlFeS$_2$ and TlFeSe$_2$, which host both localized and delocalized characters of electrons indicate the existence of delocalization of the Fe$^{3+}$ $3d$ electrons in RbFeS$_2$. The spin waves can be successfully modeled using a Heisenberg Hamiltonian and linear spin wave theory, allowing us to extract the single ion anisotropy $SD_{zz}$ and the products of the spin and the exchange interactions $SJ_{1,3}$. The simulated spectra based on the as-determined exchange parameters match well with the INS spectra. The variety of static magnetic orders yet the consistency of the spin excitations observed among several related 1D spin chain compounds highlight the interplay between local moments and delocalized electrons, resembling the situation for iron-based superconductors.

\section{ACKNOWLEDGEMENTS}

M. W. was supported by the National Natural Science Foundation of China (Grant No. 11904414, 12174454), National Key Research and Development Program of China (No. 2019YFA0705702). D. X. Y. was supported by NKRDPC-2018YFA0306001, NKRDPC-2017YFA0206203, NSFC-11974432, GBABRF-2019A1515011337, and Leading Talent Program of Guangdong Special Projects. Work at University of California,
Berkeley and Lawrence Berkeley National Laboratory was funded by the U.S. Department of
Energy, Office of Science, Office of Basic Energy Sciences, Materials Sciences and Engineering
Division under Contract No. DE-AC02-05-CH11231 within the Quantum Materials Program
(KC2202) and the Office of Basic Energy Sciences.
The experiment at Oak Ridge National Laboratory's Spallation Neutron Source was sponsored by the Scientific User Facilities Division, Office of Basic Energy Sciences, U.S. Department of Energy.

\bibliography{RbFeS2}

\end{document}